# Three-Band Anderson Lattice Model Reveals Co-Evolution of Topological and Magnetic Phases Driven by Electron Correlation

Zhong-Yi Wang[1], Ya-Min Quan[2], Yu-Xuan Sun[1], Liang-Jian Zou[2,3,*], Xiang-Long Yu[1,*]

Z.-Y. Wang, Y.-X. Sun, X.-L. Yu
[1]School of Science, Sun Yat-sen University, Shenzhen 518107, China.
*E-mail: yuxlong6@mail.sysu.edu.cn

Y.-M. Quan, L.-J. Zou
[2]Key Laboratory of Materials Physics, Institute of Solid State Physics, HFIPS, Chinese Academy of Sciences, PO Box 1129, Hefei 230031, People's Republic of China;
[3]Science Island Branch of Graduate School, University of Science and Technology of China, Hefei 230026, People's Republic of China.
*E-mail: zou@theory.issp.ac.cn

## Abstract

Understanding the interplay of band topology, strong electron correlation, and magnetic order is the fundamental core bottleneck for realizing robust high-temperature quantum anomalous Hall effect (QAHE). Conventional two-band Anderson models are limited to paramagnetic Kondo topological insulators, failing to capture coupled topological-magnetic phase evolution relevant to the QAHE benchmark $MnBi_2Te_4$ family. We develop a minimal three-band Anderson lattice model incorporating Hubbard interaction, s-d exchange coupling, and a BHZ-like topological mechanism. Using the Kotliar-Ruckenstein slave-boson approach, we map correlation-driven phase transitions at filling v=2: increasing U drives a trivial-to-Kondo topological insulator transition, then activates the third band to mediate a paramagnetic topological insulator-to-ferromagnetic metal transition. The accompanying band reconstruction—fully spin-polarized *d*-orbitals sinking below the Fermi level, leaving itinerant *p*-orbitals to dominate low-energy physics—qualitatively matches published first-principles results for $MnBi_2Te_4$. In the strong-correlation regime, exchange coupling J stabilizes a Chern-Kondo insulator (C=1) and Weyl nodal-line semimetal. Critically, we reveal full d-orbital spin polarization renders the topological gap immune to correlation-induced narrowing, resolving the long-standing strong correlation-large gap incompatibility.

Our results show excellent qualitative alignment with recent state-of-the-art QAHE experiments, providing a unified framework for correlated magnetic topological materials and new pathways to high-temperature QAHE.

**Keywords：**

Magnetic topological insulators, Anderson lattice model, Strongly correlated systems, s-d exchange model, Chern-Kondo insulator, Quantum anomalous Hall effect, $MnBi_2Te_4$

## I. INTRODUCTION

Magnetic topological insulators [1-2], as a frontier system in condensed matter physics, are fundamentally characterized by the coupling between magnetic spin confgurations and topology of electronic wavefunctions, which gives rise to a rich variety of topological quantum states. On one hand, the broken time-reversal symmetry by magnetic order provides the foundation for realizing novel topological states. These include the quantum anomalous Hall effect (QAHE) with its dissipationless chiral edge states [3-5], axion insulators exhibiting a topological magnetoelectric effect [6-10], and potential Majorana fermions in proximity to superconductors [4,11]. On the other hand, the presence of magnetic order opens avenues for the dynamic control of topological states [12-18]; for instance, altering the magnetic configuration via an external magnetic field can drive phase transitions between distinct topological phases like the axion insulator and the quantum anomalous Hall state [13,14]. These tunable topological quantum phenomena hold significant promise for applications in dissipationless electronics and topological quantum computation [4,19].

Despite numerous theoretical predictions, intrinsic magnetic topological insulator materials with ideal properties remain scarce. The emergence of $MnBi_2Te_4$ with a tetradymite-type structure has provided an ideal material platform for this field [12]. As the first confirmed intrinsically magnetic and topologically non-trivial two-dimensional van der Waals material [21,22], $MnBi_2Te_4$ offers a core advantage over earlier magnetically doped $(Bi,Sb)_2Te_3$ thin films, which realized the QAHE only at extremely low temperatures (~30 mK) [5]. It circumvents the issues of compositional inhomogeneity and magnetic disorder associated with doping, leading to superior sample quality and a larger exchange gap [20]. Consequently, the observation temperature for topological quantum effects has been significantly elevated to the liquid helium range and beyond (1 K to several tens of K) [9,22,23]. More importantly, this system exhibits multidimensional tunability: the layer-number parity modulates its electronic ground state—odd-layer films tend to exhibit the QAHE, while even-layer ones favor the axion insulator state [13,22,24]; an external magnetic field can drive the bulk phase from an A-type antiferromagnetic topological insulator to a ferromagnetic Weyl semimetal [10,12,14]; and its chemical composition is richly variable, forming an extensive $MB_2T_4$ (M = Mn or other transition/rare-earth metals, B = Bi/Sb, T = Te/Se/S) material family[12,25-28]. Furthermore, by modifying the stacking periodicity and stoichiometry, a series of related materials such as $Mn_2Bi_2Te_5$ [29], $MnBi_8Te_{13}$ [30], $MnBi_6Te_{10}$ [31], and $MnBi_4Te_7$ [32] can be derived. These properties collectively establish the $MnBi_2Te_4$ family as an ideal platform for realizing and thoroughly investigating various topological quantum states.

However, a core challenge persists: the observation temperature for most topological quantum effects remains far below the liquid nitrogen temperature, severely hindering practical applications. The discovery of more magnetic topological materials with high characteristic temperatures is thus imperative. Yet, the interplay of magnetism and topology makes their theoretical prediction more difficult than for non-magnetic systems. At the level of computational materials science, the presence of strong correlation effects poses challenges for first-principles methods based on the single-electron approximation to accurately predict their electronic structures [33]. Additionally, the introduction of magnetic degrees of freedom, combined with the crystal space group, results in a vast number of magnetic space groups [17]. While this enhances the diversity of magnetic topological states, it also increases the complexity of research. In terms of theoretical modeling, descriptions of nontrivial band topology using the Kane-Mele, Haldane, or Bernevig-Hughes-Zhang (BHZ) models [34-38], when augmented with Hubbard interactions, have largely focused on the half-filling scenario. These studies primarily address the phase transition from a topological insulator at small U to a trivial Mott insulator at large U [34,35]. Systematic investigations into how magnetism specifically regulates topological properties, as well as the physical picture under conditions deviating from half-filling, are still lacking[37].

Therefore, this study aims to explore the universal regulatory mechanisms of strong correlation effects and their resulting magnetism on the band structure of topological materials. To this end, drawing inspiration from the key [Te–Mn–Te] trilayer Hamiltonian in $MnBi_2Te_4$ [39], which is identified as a core carrier of both topology and magnetism, we adopt and develop a minimal three-band low-energy effective model. It captures the essential features of the low-energy electronic structure and the crystalline symmetry (point group $D_{3d}$) of such materials, while abstracting away their complex details to reveal physics that transcends any specific compound. The core of the model lies in its integration of a topological origin mechanism analogous to that in the BHZ model [40,41], the Hubbard interaction (U) describing electron correlations, and the s-d exchange coupling (J). Consequently, it couples the three core degrees of freedom—topology, correlation, and magnetism—within a unified framework. We employ this model to systematically study their interplay, focusing on the electronic fillings $\nu = 1$ and $\nu = 2$.

The remainder of this paper is organized as follows. In **Sec. II**, we detail the construction of this model. The theoretical framework is based on the periodic Anderson model [42], and strong correlation effects are treated using the Kotliar-Ruckenstein slave-boson mean-field approach [43,44]. **Sec. III** presents the main results. We first systematically analyze how the correlation strength U drives magnetic phase transitions and topological evolution under both paramagnetic and ferromagnetic conditions for systems at fillings $\nu = 2$ and $\nu = 1$, respectively. Notably, in contrast to previous two-band correlated models [45-47], the three-band structure of our study reveals that the involvement of the third band plays a crucial role in driving the transition to a ferromagnetic metallic phase under strong correlation, building upon a correlation-induced Kondo topological insulator phase [48]. Subsequently, we focus on the strong correlation limit (large U), elucidating the decisive role of the exchange coupling J in regulating the topological bands and revealing its key mechanism in inducing topological states such as the Chern-Kondo insulator [49] (whose unique prospect for a large gap will be discussed in detail). Finally, **Sec. IV** summarizes the principal findings of this work. We emphasize the discussion on magnetic regulation by correlations (e.g., enhanced ferromagnetic stability), the systematic evolution of the band structure under correlation, and the distinctive properties of the Chern-Kondo insulator phase. The section concludes with an outlook on future research directions for exploring the QAHE at high characteristic temperatures in correlated materials.

## II. EFFECTIVE MODEL AND MEAN-FIELD APPROACH

To investigate the regulatory mechanism of electron correlations in magnetic topological materials, this work constructs a minimal low-energy effective model designed to simultaneously describe both the magnetic and topological properties of the system. The model is inspired by the low-energy electronic structure revealed in first-principles (DFT) calculations for monolayer MnBi$_2$Te$_4$ performed without correlation (U = 0). Specifically, the DFT-computed density of states (Fig. 1(a)) indicates a dominant Mn-3d orbital character around the energy zero (corresponding to the Fermi level E$_F$), while the states at positive energies just above E$_F$ are primarily derived from Bi/Te-p orbitals. In contrast, previous DFT studies [12] have shown that with the inclusion of an electron interaction (U = 4 eV), the *d*-orbitals undergo significant splitting, provide the dominant magnetic moment, and shift away from the E$_F$, consequently yielding a low-energy region dominated by *p*-orbitals. To capture the essential low-energy physics of such materials while preserving the lattice symmetry of MnBi$_2$Te$_4$, we focus on the [Te–Mn–Te] trilayer (Fig. 1(b)), identified in Ref.[39] as the key structural unit governing both magnetism and topology. This structure features a tripled triangular lattice stacking, where the lattice sites of each layer are located at the centers of the adjacent layers, thereby maintaining the same point group symmetry (D$_{3d}$) as the parent material. To construct the minimal model, we extract the essential low-energy degrees of freedom from this unit. At the Mn site in the middle layer, the local magnetic moment and correlation effects are described by introducing a single effective *d* orbital. At each of the upper and lower layer sites—formally corresponding to the Te positions—a single effective "*p*" orbital is introduced, physically interpreted as a composite degree of freedom representing the Bi/Te-*p* orbitals, which constitute the low-energy conduction states as suggested by the DOS. This design condenses the complex band structure of the actual material into a minimal three-band model, which facilitates the clarification of the universal physical principles underlying the interplay between magnetism and topology.

The specific selection of the orbital basis follows the Wannier Hamiltonian construction scheme for the [Te-Mn-Te] system described in the existing Ref. [39], but assigns a new physical interpretation to the Te sites:

**"Mn" *d*-orbital basis:** Under the influence of the triangular crystal field (point group D$_{3d}$) and spin-orbit coupling (SOC), the originally five-fold degenerate 3*d* orbitals of the Mn$^{2+}$ (3*d*$^5$) ion split. To construct the low-energy effective model, we neglect other energy states far from the Fermi level and select only the Kramers doublet closest to it, namely the states with total angular momentum quantum number j = 5/2 and projection $m_j$ = ±5/2, denoted as $|d_{5/2, \pm 5/2}\rangle$. This state can be explicitly expressed as a direct product of an orbital state with orbital angular momentum projection $l_z = \pm 2$ (corresponding to the $|d_{x2-y2} \pm id_{xy}\rangle$ orbital) and a spin state with projection $s_z = \pm 1/2$, effectively describing the local magnetism and correlation effects at the Mn site.

**"Te" p-orbital basis:** For the atoms in the upper and lower layers, the $p$-orbitals split under the triangular crystal field (point group $C_{3v}$) and SOC into three Kramers doublets. To simplify the model, we select one of these doublets, specifically the state with j = 3/2, $m_j = \pm 3/2$, denoted as $|p_{3/2, \pm 3/2}\rangle$, as our basis. This state corresponds to the direct product state of an orbital with angular momentum projection $l_z = \pm 1$ (the $|p_x \pm ip_y\rangle$ orbital) and spin projection $s_z = \pm 1/2$. Here, "Te" should be understood as representing the composite $p$-orbital contribution from Bi/Te atoms, describing the conduction bands above the Fermi level in the DFT results, differing in physical connotation from the original Ref. [39].

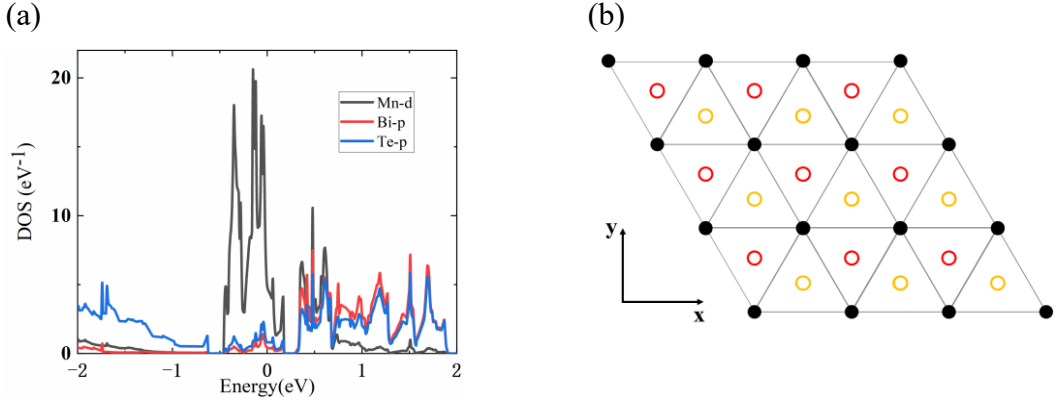

Fig. 1. **Electronic structure at U=0 and the core [Te-Mn-Te] structural unit of monolayer MnBi$_2$Te$_4$.** (a) Projected density of states (DOS) from DFT calculations (U=0), showing the orbital contributions of different elements. The energy zero corresponds to the Fermi level. (b) Schematic of the [Te-Mn-Te] sandwich structure: Mn atoms (solid circles) form a triangular lattice in the middle layer; Te atoms (triangle vertices) are located above and below, also forming triangular lattices. The red open circles mark the projections of the upper-layer Te atoms onto the middle Mn plane, while the yellow ones represent those from the lower layer.

Based on the above basis selection, the tight-binding Hamiltonian $\widehat{H}_{TB}(k)$ for a single electron of the system can be constructed from the following spin orbitals states: the $|d_{5/2,\pm 5/2}\rangle$ states at the "Mn" site, and the $|1, p_{3/2,\pm 3/2}\rangle$ and $|2, p_{3/2,\pm 3/2}\rangle$ states at the upper and lower "Te" sites, respectively. The matrix elements of the Hamiltonian are determined by six real parameters and their corresponding momentum-dependent functions: $\varepsilon_p$ and $\varepsilon_d$ denote the on-site energies of the $d$ and $p$ orbitals, respectively; $t_d$ and $t_p$ describe the nearest-neighbor hopping between *d-d* and *p-p* orbitals within the same atomic layer; $t_{pd}$ characterizes the *p-d* orbital hybridization hopping between adjacent "Mn" and "Te" layers; and $t_{pp}$ describes the *p-p* orbital hopping between the two "Te" layers. Since the spin-orbit coupling effect is implicitly included in the construction of the basis states [41], the Hamiltonian is block-diagonal in spin space and can be written as:

$$\hat{H}_{TB}(k) = \begin{pmatrix} \hat{H}_+(k) & 0 \\ 0 & \hat{H}_-(k) \end{pmatrix}, \tag{1}$$

here $\hat{H}_\pm(\mathbf{k})$ correspond to the spin-up (+) and spin-down (−) subspaces, respectively. To satisfy time-reversal symmetry, $\hat{H}_-(\mathbf{k}) = \hat{H}_+^*(-\mathbf{k})$. The matrix form of $\hat{H}_\pm(\mathbf{k})$ is:

| $\hat{H}_\pm(k)$ | $\|1, p_{3/2, \pm 3/2}\rangle$ | $\|d_{5/2, \pm 5/2}\rangle$ | $\|2, p_{3/2, \pm 3/2}\rangle$ |
|---|---|---|---|
| $\langle 1, p_{3/2, \pm 3/2}\|$ | $\varepsilon_p - t_p G(\mathbf{k})$ | $-t_{pd} F_{\pm 1}^*(\mathbf{k})$ | $-t_{pp} F_0(\mathbf{k})$ |
| $\langle d_{5/2, \pm 5/2}\|$ | $-t_{pd} F_{\pm 1}(\mathbf{k})$ | $\varepsilon_d + t_d G(\mathbf{k})$ | $t_{pd} F_{\mp 1}^*(\mathbf{k})$ |
| $\langle 2, p_{3/2, \pm 3/2}\|$ | $t_{pp} F_0^*(\mathbf{k})$ | $t_{pd} F_{\mp 1}(\mathbf{k})$ | $\varepsilon_p - t_p G(\mathbf{k})$ |

(2)

Here, $G(\mathbf{k})$ and $F_{0,\pm 1}(\mathbf{k})$ are geometric phase factors that depend on the lattice structure. Their explicit forms are:

$$G(\mathbf{k}) = 2[\cos(ak_x) + \cos(ak_x/2 + \sqrt{3}ak_y/2) + \cos(ak_x/2 - \sqrt{3}ak_y/2)]$$

$$F_n(\mathbf{k}) = e^{-i\mathbf{k}\cdot\boldsymbol{\tau}_a} + e^{\frac{i2\pi n}{3}} e^{-i\mathbf{k}\cdot\boldsymbol{\tau}_b} + e^{-\frac{i2\pi n}{3}} e^{-i\mathbf{k}\cdot\boldsymbol{\tau}_c}, n = 0, \pm 1 \tag{3}$$

where $\boldsymbol{\tau}_a = (0, \sqrt{3}/3)a$, $\boldsymbol{\tau}_b = (1/2, -\sqrt{3}/6)a$, and $\boldsymbol{\tau}_c = (-1/2, -\sqrt{3}/6)a$ are the position vectors (with $a$ being the lattice constant of the triangular lattice) of the three equivalent nearest-neighbor "Te" atoms in the first layer relative to the "Mn" atom at the origin.

It is crucial to emphasize that the key ingredient for generating nontrivial topological band structure in this model originates from the interlayer *p-d* orbital hybridization term $F_{\pm 1}(\mathbf{k})$ [39]. Constrained by the threefold rotational symmetry ($C_3$) of the system along the z-axis, the hopping integrals between orbitals with different angular momentum quantum numbers exhibit directional dependence. Taking the hopping between Mn and a "Te" atom in the spin-up channel as an example, the hopping integrals along the three nearest-neighbor directions satisfy:

$$t_{pd} = t_{pd}(\boldsymbol{\tau}_a) = e^{-i2\pi/3} t_{pd}(\boldsymbol{\tau}_b) = e^{-i4\pi/3} t_{pd}(\boldsymbol{\tau}_c) \tag{4}$$

This relation introduces additional direction-dependent phase factors $e^{\pm i2\pi/3}$ into $F_{\pm 1}(\mathbf{k})$. Consequently, the *p-d* hybridization exhibits distinct momentum-space dependencies in the spin-up and spin-down subsystems. This mechanism is essential for inducing the nontrivial topological band structure and is analogous to the origin of topology in the BHZ model [39-41]. In contrast, the intralayer hopping described by $G(\mathbf{k})$ and the direct hopping between the two "Te" layers via $F_0(\mathbf{k})$ occur between orbitals with the same angular momentum projection and thus lack such directional-dependent phase characteristics.

After constructing the tight-binding model, we can formulate the system's Hamiltonian in second quantization. We define the momentum-space creation (annihilation) operators: $C_{1k\sigma}^\dagger(C_{1k\sigma})$ and $C_{2k\sigma}^\dagger(C_{2k\sigma})$ for the upper and lower "Te" sites, respectively, and $d_{k\sigma}^\dagger(d_{k\sigma})$ for the "Mn" site, where $\sigma = \pm$ denotes the spin. The non-interacting Hamiltonian is then given by:

$$H_0 = \sum_{k\sigma} \hat{\psi}_{k\sigma}^\dagger [\hat{H}_\sigma(\mathbf{k}) - \mu \hat{I}] \hat{\psi}_{k\sigma} \tag{5}$$

where $\mu$ is the chemical potential, $\hat{I}$ is the identity matrix, and $\hat{\psi}_{k\sigma}^\dagger = (C_{1k\sigma}^\dagger, d_{k\sigma}^\dagger, C_{2k\sigma}^\dagger)$. To incorporate electron correlation effects, an on-site density-density interaction is introduced on the *d*-orbitals:

$$H_U = U/2 \sum_i \sum_{\sigma=\pm} d_{i\sigma}^\dagger d_{i\sigma} d_{i\bar\sigma}^\dagger d_{i\bar\sigma} \tag{6}$$

Here, $i$ is the lattice site index in real space. This model is essentially a periodic Anderson model (PAM), containing localized *d* electrons, conduction *p* electrons, and the hybridization between them. To effectively describe the coupling between the strongly localized *d* electrons with large magnetic moments (in the large-U limit) and the spins of the conduction *p* electrons, we additionally introduce an s-d-type exchange interaction term [50]:

$$H_{s-d} = -\frac{J}{N} \sum_{\lambda=1}^2 \sum_{i,\delta_\lambda} \sum_{k,k'} \sum_{\sigma,\sigma'} e^{i(\mathbf{k}'-\mathbf{k})\cdot(\mathbf{i}+\delta_\lambda)} \hat{\mathbf{S}}_i \cdot \hat{\boldsymbol{\tau}}_{\sigma\sigma'} C_{\lambda k\sigma}^\dagger C_{\lambda k'\sigma'} \tag{7}$$

Here, $J$ is the exchange coupling constant; $\boldsymbol{\delta}_1$ and $\boldsymbol{\delta}_2$ denote the vectors to the three nearest-neighbor *p*-orbital lattice points from the magnetic moment at site $i$ on the upper and lower layers, respectively; $\hat{\mathbf{S}}_i = \frac{1}{2} \sum_{\sigma,\sigma'} d_{i\sigma}^\dagger \hat{\boldsymbol{\tau}}_{\sigma\sigma'} d_{i\sigma'}$ is the local spin operator; and $\hat{\boldsymbol{\tau}}$ is the vector of Pauli matrices.

It is important to note that this term is not a standard component of the PAM but rather an effective interaction derived in the Kondo limit via the Schrieffer-Wolff transformation [51,52]. Its purpose is to more accurately capture the coupling physics between the localized magnetic moments (resulting from strongly localized *d*-electrons in the large-U limit) and the conduction electron spins. Therefore, this term is considered only when the system is in the large-U limit and exhibits substantial local moments.

To simplify the analysis and focus on the core physics, in this study we consider the following scenario: the magnetization of the *d*-orbitals is aligned along the z-direction, and we assume the scattering of conduction electrons by the local magnetic moments is an elastic process that conserves momentum (i.e., $\mathbf{k}' = \mathbf{k}$). Under these conditions, the $H_{s-d}$ term simplifies to an effective magnetic field acting on the conduction p electrons from the *d*-orbital magnetic moment:

$$H_{s-d} = -\frac{3J}{N}\sum_{\lambda=1}^{2}\sum_{i}\sum_{k}\sum_{\sigma=\pm}\frac{\sigma}{2}(d_{i+}^{\dagger}d_{i+} - d_{i-}^{\dagger}d_{i-})C_{\lambda k\sigma}^{\dagger}C_{\lambda k\sigma} \quad (8)$$

The total Hamiltonian of the system is:

$$H = H_0 + H_U + H_{s-d} \quad (9)$$

To treat the strong correlation effects, we employ the Kotliar–Ruckenstein (K–R) slave-boson mean-field approach [42-44], rewriting the d-electron operator as:

$$d_{i\sigma} \to f_{i\sigma}z_{i\sigma} \quad (10)$$

where $z_{i\sigma}$ is a bosonic projection operator:

$$z_{i\sigma} = \frac{e_{i}^{\dagger}p_{i\sigma} + p_{i\bar{\sigma}}^{\dagger}b_{i}}{\sqrt{(1-b_{i}^{\dagger}b_{i} - p_{i\sigma}^{\dagger}p_{i\sigma})(1-e_{i}^{\dagger}e_{i} - p_{i\bar{\sigma}}^{\dagger}p_{i\bar{\sigma}})}} \quad (11)$$

The bosonic operators— $e_i^{\dagger}$ (empty), $p_{i\sigma}^{\dagger}$ (single occupation with spin $\sigma$), and $b_i^{\dagger}$ (double occupation)—satisfy the constraint conditions:

$$e_i^{\dagger}e_i + \sum_{\sigma}p_{i\sigma}^{\dagger}p_{i\sigma} + b_i^{\dagger}b_i = 1; \quad p_{i\sigma}^{\dagger}p_{i\sigma} + b_i^{\dagger}b_i = f_{i\sigma}^{\dagger}f_{i\sigma}. \quad (12)$$

Consequently, the d-operator products in the interaction terms are replaced as:

$$\frac{1}{2}\sum_{\sigma=\pm}d_{i\sigma}^{\dagger}d_{i\sigma}d_{i\bar{\sigma}}^{\dagger}d_{i\bar{\sigma}} \to b_i^{\dagger}b_i; \quad d_{i+}^{\dagger}d_{i+} - d_{i-}^{\dagger}d_{i-} \to p_{i+}^{\dagger}p_{i+} - p_{i-}^{\dagger}p_{i-}. \quad (13)$$

Applying the saddle-point approximation to the bosonic fields, we assume their expectation values are uniform (independent of lattice site *i*) and replace the bosonic operators with static mean values $e, b, p_{\sigma}$. Lagrange multipliers $\lambda_{\sigma}$ and $\lambda'$, along with the particle filling $\nu$, are introduced to enforce the constraints. This yields the mean-field Hamiltonian:

$$H_{MF} = H_{Bose} + H_{Fermi} \quad (14)$$

where

$$H_{Bose} = NUb^2 + N\mu\nu - N\sum_{\sigma}\lambda_{\sigma}(p_{\sigma}^2 + b^2) + N\lambda'(e^2 + \sum_{\sigma}p_{\sigma}^2 + b^2 - 1),$$

$$H_{Fermi} = \sum_{k\sigma=\pm}\hat{\varphi}_{k\sigma}^{\dagger}[\hat{H}_{\sigma}(k, z_{\sigma}, \lambda_{\sigma}) - \mu\hat{I}]\hat{\varphi}_{k\sigma}. \quad (15)$$

Here, $\hat{\varphi}_{k\sigma}^{\dagger} = (C_{1k\sigma}^{\dagger}, f_{k\sigma}^{\dagger}, C_{2k\sigma}^{\dagger})$, and the renormalized Bloch Hamiltonian matrix is:

$$\hat{H}_{\sigma}(k, z_{\sigma}, \lambda_{\sigma})$$
$$= \begin{pmatrix} \varepsilon_p - t_p G(\boldsymbol{k}) - 3J<S_z>\sigma & -t_{pd}\bar{z}_\sigma F_\sigma^*(\boldsymbol{k}) & -t_{pp}F_0(\boldsymbol{k}) \\ -t_{pd}\bar{z}_\sigma F_\sigma(\boldsymbol{k}) & \varepsilon_d + t_d \bar{z}_\sigma^2 G(\boldsymbol{k}) + \lambda_\sigma & t_{pd}\bar{z}_\sigma F_{\bar{\sigma}}^*(\boldsymbol{k}) \\ t_{pp}F_0^*(\boldsymbol{k}) & t_{pd}\bar{z}_\sigma F_{\bar{\sigma}}(\boldsymbol{k}) & \varepsilon_p - t_p G(\boldsymbol{k}) - 3J<S_z>\sigma \end{pmatrix}$$

(16)

In the above expression, $\langle \hat{S}_i^z \rangle = \frac{1}{2}(p_+^2 - p_-^2)$ represents the average local magnetic moment on the $d$-orbital, and $\bar{z}_\sigma = (e^2 + p_\sigma^2)^{-1/2}(ep_\sigma + p_{\bar{\sigma}}b)(b^2 + p_\sigma^2)^{-1/2}$ is the band renormalization factor. Here, the notation $F_\pm(\boldsymbol{k})$ is a shorthand for the geometric factor $F_{\pm 1}(\boldsymbol{k})$ introduced earlier, and the Lagrange multiplier $\lambda_\sigma$ characterizes the effective energy shift of the $d$-orbital. Finally, the ground state and electronic structure of the system are determined by solving the self-consistent equations obtained from $\langle \partial H_{\mathrm{MF}} / \partial \theta \rangle = 0$, where $\theta = \{e, p_\sigma, b, \lambda_\sigma, \lambda', \mu\}$.

Simultaneously, to align with the low-energy picture revealed by DFT, we predefine the on-site energy of the $d$-orbital, $\varepsilon_d$, to be lower than that of the $p$-orbital, $\varepsilon_p$, i.e., we set a positive energy difference $\varepsilon_g = \varepsilon_p - \varepsilon_d > 0$. This setting aims to qualitatively reflect the characteristics of materials like $MnBi_2Te_4$ in the non-correlated (U=0) scenario: the bands dominated by Mn-3$d$ orbitals reside near the Fermi level ($E_F$), while the conduction bands immediately above $E_F$ are contributed by Bi/Te-$p$ orbitals.

Secondly, the filling factor $\nu$ in our model is defined as the total number of electrons occupying the aforementioned low-energy effective orbital basis. DFT calculation results indicate that the Fermi level of the system lies within the Mn-3$d$ dominated bands. Therefore, we focus on the filling scenarios of $\nu = 1$ and $\nu = 2$. These two cases correspond to electronic occupation states near the Fermi surface being primarily contributed by $d$-orbitals, making them an ideal regime for investigating how correlation effects (U) regulate the magnetism in this region and its interplay with the topology of the neighboring $p$-orbital bands. In particular, considering that the $Mn^{2+}$ ion (3$d^5$ electron configuration) in $MnBi_2Te_4$ is in a half-filled state, the $\nu = 1$ filling scenario qualitatively simulates the low-energy electronic structure of $MnBi_2Te_4$ in its intrinsic condition.

Furthermore, this study primarily focuses on the correlated topological effects under paramagnetic and out-of-plane ferromagnetic orders. This choice is based on the material foundation of the model: it is a low-energy abstraction of monolayer $MnBi_2Te_4$ and its homologous materials. In such materials, the intra-layer magnetic ions are connected via ligands (e.g., Te) with bond angles close to 90° (e.g., the Mn-Te-Mn bond angle is approximately 86°) [13]. According to the Goodenough-Kanamori rules, the superexchange interaction at this bond angle intrinsically favors intra-layer ferromagnetic coupling [12]. This has also been validated in two-dimensional ferromagnets with similar structures, such as CrI$_3$ and Cr$_2$Ge$_2$Te$_6$ [53,54]. Therefore, setting ferromagnetic order as the magnetic background for the model is justified.

Finally, since there is no coupling between the upper and lower spin subspaces of the system, when judging the band topology in the paramagnetic phase, we can still directly analyze it through changes in the spin Chern number for each subspace [55], without relying on the $Z_2$ topological invariant.

Based on the aforementioned model framework and parameter settings, we proceed to discuss the calculation results.

## III. RESULTS AND DISCUSSION

In this work, by tuning the on-site energy difference between the *p* and *d* orbitals ($\varepsilon_g = \varepsilon_p - \varepsilon_d$), we systematically investigate the influence of electron correlation strength U on the system's states. The values of the hopping integral parameters are partially referenced from data reported for the [Te–Mn–Te] system (Ref. [39]) and are set as: $t_d = 0.15$, $t_p = 0.25$, $t_{pd} = 0.1$, and $t_{pp} = 0.2$, all in units of eV.

Fig. 2 illustrates the influence of the *p-d* orbital energy level difference $\varepsilon_g$ on the band structure in the absence of electronic correlation (U = 0). When $\varepsilon_g$ = 3.3 eV, the system behaves as a trivial insulator. As $\varepsilon_g$ decreases, the bands corresponding to the *d* and *p* orbitals begin to overlap, with the *p-d* hybridization opening a band gap accompanied by band inversion. Calculations of the spin Chern number confirm that the system enters a topological insulator phase with a Chern number of 1 at $\varepsilon_g$ = 2.3 eV. Continuing to reduce $\varepsilon_g$ to 0.7 eV, the *p*-orbital band which originally resided at a higher energy level (i.e., the highest-energy band among the three) begins to cross the Fermi level, placing the system near the critical point of a metal-insulator transition. Finally, at $\varepsilon_g$ = 0.3 eV, the system becomes fully metallic. Building on this, we further examine the effects of electron correlation U and the s-d exchange interaction J on the system under both paramagnetic (PM) and ferromagnetic (FM) states for different $\varepsilon_g$ values.

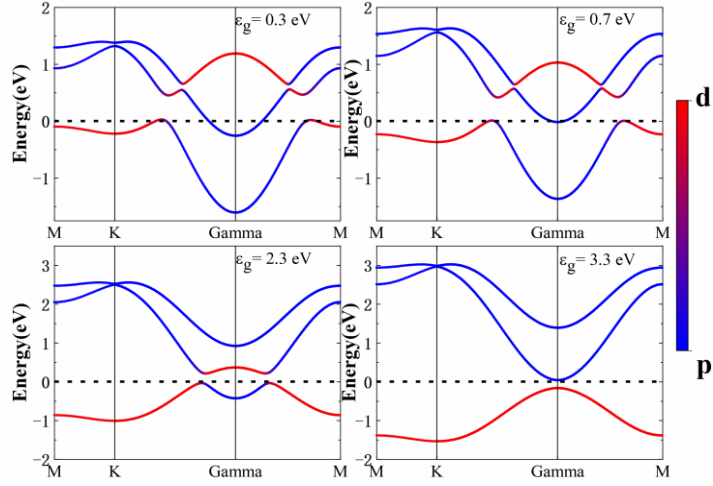

**Fig. 2. Band structures under different on-site energy differences $\varepsilon_g$ at U = 0 (filling factor ν = 2).** The red and blue portions of the bands indicate dominance by *d*-orbital and *p*-orbital components, respectively. Decreasing $\varepsilon_g$ values correspond to an increasing overlap between the p and d orbitals, leading the system to successively exhibit a trivial insulator, a topological insulator, a critical insulator-metal state, and a metallic state.

## A. Correlation effects at J = 0 (pure periodic Anderson model) and filling ν = 2

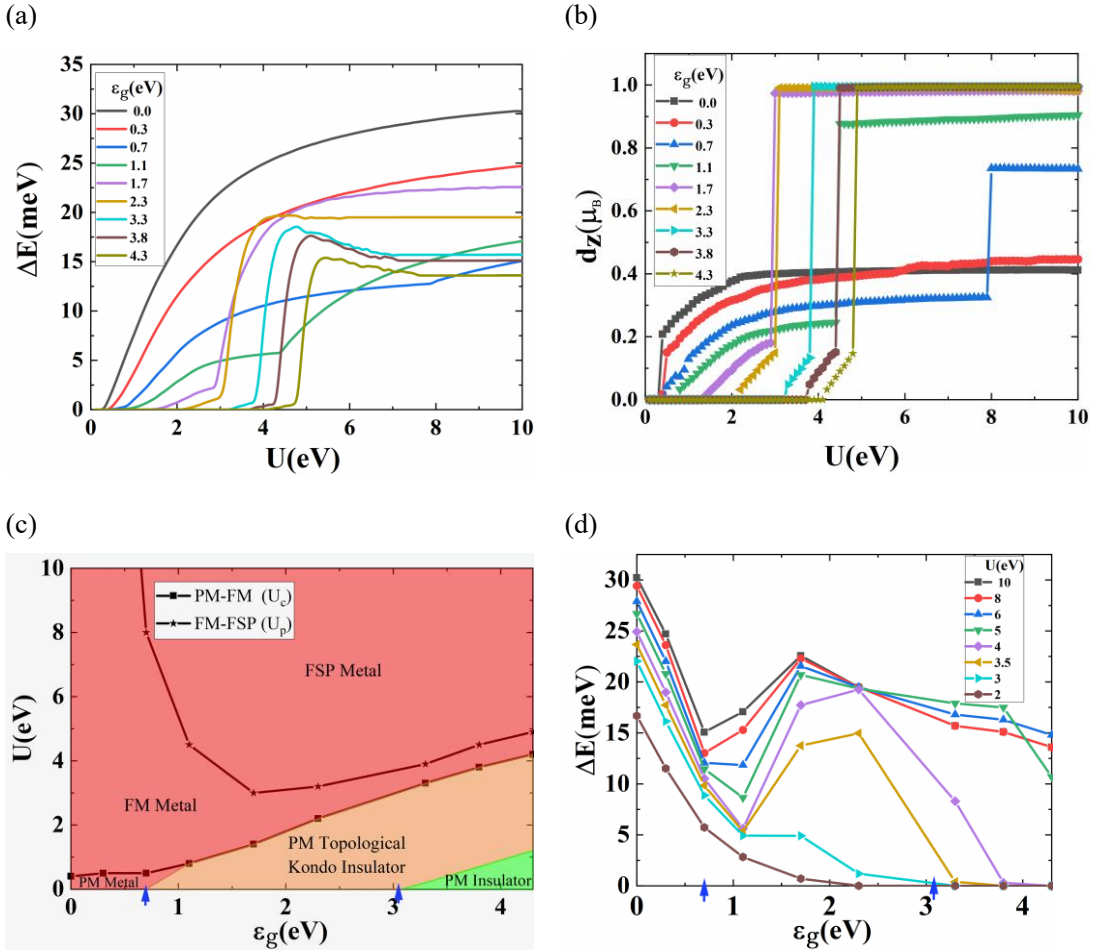

**Fig. 3. Ground-state energy, magnetism, and electronic phases of the system at filling ν = 2 tuned by the electron correlation strength U.** (a) Energy difference between the PM and FM states, ΔE = $E_{PM}$ − $E_{FM}$, as a function of U for different $\varepsilon_g$. (b) Corresponding evolution of the *d*-orbital magnetic moment with U. (c) U-$\varepsilon_g$ phase diagram of the system. The red, orange, and green areas represent the metallic phase, topological insulator phase, and trivial insulator phase, respectively. The square-dotted line and the star-dotted line indicate the critical values $U_c$ (PM-FM phase boundary) and $U_p$ (FM-FSP phase boundary), respectively. (d) ΔE as a function of $\varepsilon_g$ for different values of U, derived from the data in (a). In (c) and (d), the blue arrows on the horizontal axes mark two characteristic points at U = 0: $\varepsilon_g$ = 0.7 eV (metal-insulator crossover) and $\varepsilon_g$ = 3.08 eV (topological critical point).

Figs. 3(a) and 3(b) show the evolution of the system energy difference (ΔE = $E_{PM}$ − $E_{FM}$) and the *d*-orbital magnetic moment with U for different $\varepsilon_g$. As seen in Fig. 3(a), regardless of the initial metallic or insulating phase, the energy of the PM state eventually becomes higher than that of the FM state with increasing U, signifying the system's entry into the FM phase. This corresponds to the emergence of a finite magnetic moment at a critical value $U_c$ in Fig. 3(b). The magnetic moment then increases monotonically with further increase in U. When U increases to another critical value $U_p$, an inflection point appears in the energy curve, accompanied by a jump in the magnetic moment to saturation, indicating a phase transition from a partially spin-polarized FM state to a fully spin-polarized (FSP) state. The saturated magnetic moment decreases with decreasing $\varepsilon_g$ (i.e., enhanced *d-p* orbital overlap), with a particularly pronounced reduction in the region where $\varepsilon_g$ < 1.7 eV (corresponding to states closer to metallic at U = 0). This can be attributed to the transfer of *d*-orbital electrons to the *p*-orbitals.

Based on these results, we construct the U-$\varepsilon_g$ phase diagram of the system (Fig. 3c). It can be seen that as $\varepsilon_g$ decreases (*d-p* orbital overlap strengthens), the critical value $U_c$ for the PM-FM transition gradually decreases. Meanwhile, the critical value $U_p$ for the FM-FSP transition exhibits non-monotonic behavior with varying $\varepsilon_g$: in the range $\varepsilon_g$ > 1.7 eV (corresponding to the insulating region at U = 0), $U_p$ decreases with decreasing $\varepsilon_g$; whereas in the range $\varepsilon_g$ < 1.7 eV (near the metallic region at U = 0), $U_p$ increases with decreasing $\varepsilon_g$. When $\varepsilon_g$ is further reduced below 0.7 eV (metallic region at U = 0), the $U_p$ value exceeds the computational range of this study (0 – 10 eV).

Further analysis of the spectral evolution under correlation effects (representative results are shown in Appendix Fig. S1 and S2) indicates that for systems with $\varepsilon_g > 0.7$ eV (insulating region at U = 0), the originally unoccupied, highest-energy third band begins to cross the Fermi level with increasing U, thereby driving the system into a metallic phase. Notably, the band structures near the PM-FM transition points are precisely located within the critical region of the insulator-metal transition. Consequently, the PM-FM phase boundary in Fig. 3(c) essentially coincides with the insulator-metal boundary. In particular, for systems with $\varepsilon_g > 3.08$ eV (trivial insulator at U = 0), band inversion occurs with increasing U (Fig. S2), transforming the system from an initial trivial insulator into a topological insulator. This transition originates from the correlation-induced shift of the *d*-orbital energy levels, which enhances the *d*-*p* band overlap, constituting a behavior akin to that of a topological Kondo insulator [45,46]. However, at larger *U* values, the involvement of the third band disrupts this PM topological Kondo insulator phase, causing it to give way to a FM metal.

To investigate the stability of the FM phase, we systematically analyzed the variation of ΔE with $\varepsilon_g$ for different U values (Fig. 3(d)). Here, the terms "insulating region" and "metallic region" specifically refer to the nature of the bands as determined by $\varepsilon_g$ under the condition of U = 0. Under weaker correlation conditions (U < 3 eV), ΔE increases monotonically with decreasing $\varepsilon_g$. Under stronger correlation conditions (U ≥ 4 eV), ΔE still increases with decreasing $\varepsilon_g$ in the insulating region ($\varepsilon_g \approx 4.3 \sim 2$ eV), turns to decrease in the region approaching metallicity ($\varepsilon_g \approx 2 \sim 1$ eV), and rises again near and upon entering the metallic region ($\varepsilon_g < 1$ eV).

**B. Correlation effects at J = 0 (pure periodic Anderson model) and filling ν = 1**

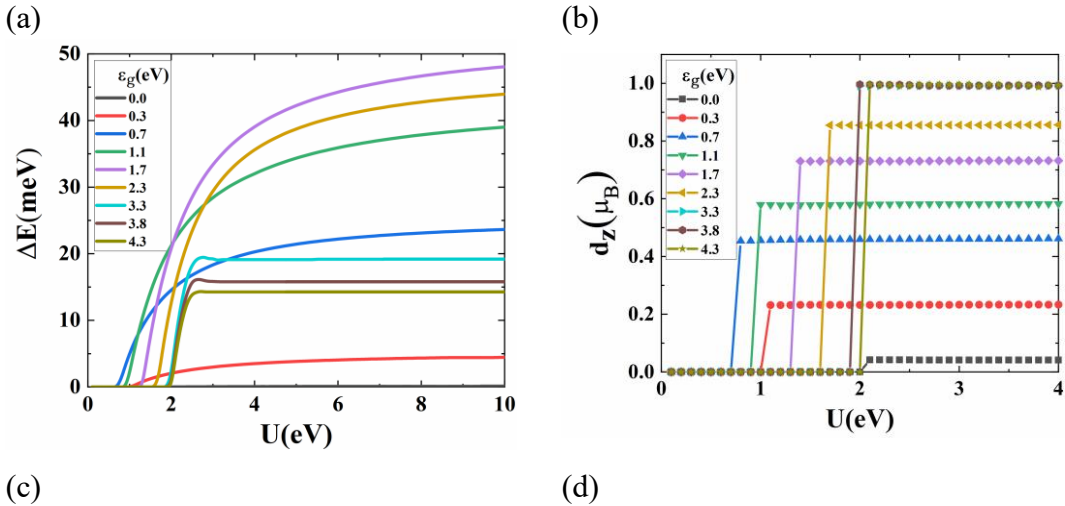

(a) (b)

(c) (d)

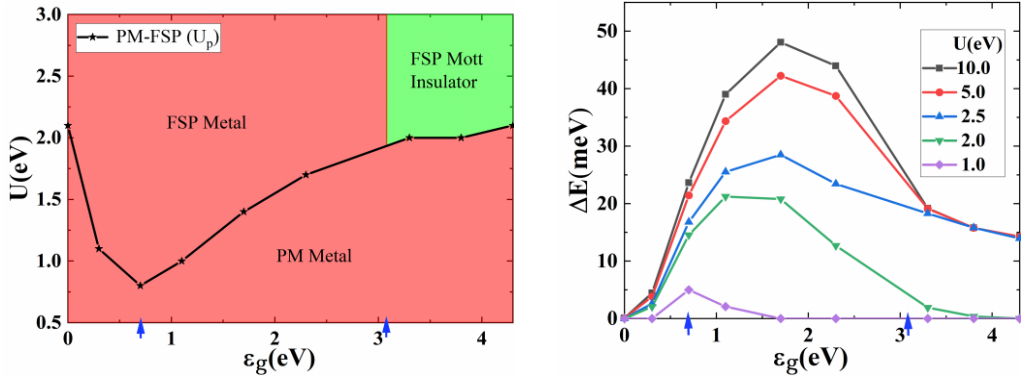

**Fig. 4. Ground-state energy, magnetism, and electronic phases of the system at filling ν = 1 tuned by the electron correlation strength U.** (a) Energy difference between the paramagnetic (PM) and ferromagnetic (FM) states, ΔE = $E_{PM}$ – $E_{FM}$, as a function of U for different $\varepsilon_g$. (b) Corresponding evolution of the *d*-orbital magnetic moment with U. (c) U-$\varepsilon_g$ phase diagram of the system. The red and green areas represent the metallic and insulating phases, respectively. The stippled line indicates the phase boundary from the PM to the FSP phase, with the critical value denoted as $U_p$. (d) ΔE as a function of $\varepsilon_g$ for different values of U. In (c) and (d), the blue arrows on the horizontal axes mark two characteristic points of the bands at U = 0: $\varepsilon_g$ = 0.7 eV (metal-insulator crossover) and $\varepsilon_g$ = 3.08 eV (topological critical point).

Figures 4(a) and 4(b) show the evolution of the system energy and the *d*-orbital magnetic moment with the correlation strength U for different $\varepsilon_g$. Similar to the case of ν = 2, as U increases, the energy of the PM state eventually surpasses that of the FM state, marking the system's entry into the FM phase. However, there is a significant difference in the evolution of the magnetic moment: as shown in Fig. 4(b), upon entering the FM phase, the magnetic moment jumps directly from zero to its saturation value, indicating that only a single phase transition from PM to a FSP state occurs here. The saturated magnetic moment value also decreases with decreasing $\varepsilon_g$, and due to the lower filling, this decreasing trend is more rapid and pronounced; the moment nearly vanishes as $\varepsilon_g$ approaches zero.

Based on these results, we construct the corresponding U-$\varepsilon_g$ phase diagram (Fig. 4(c)). It can be seen that, akin to the ν =2 case, the critical value $U_p$ for the PM-FSP transition exhibits non-monotonic behavior with decreasing $\varepsilon_g$: in the range $\varepsilon_g$ = 4.3 ~ 0.7 eV, $U_p$ decreases with decreasing $\varepsilon_g$; when $\varepsilon_g$ < 0.7 eV, $U_p$ instead increases with decreasing $\varepsilon_g$. A comparison with the results for ν = 2 (Fig. 3(c)) further reveals that the U value required for the system to reach magnetic saturation is generally smaller for ν = 1.

Further analysis of the spectral evolution under correlation effects (see Appendix Fig. S3 and S4) indicates that due to ν = 1 and the charge transfer induced by *p-d* band overlap and hybridization, the system exhibits a metallic state over the vast majority of the parameter space. However, for systems in the *p-d* orbital non-overlap region at U = 0 ($\varepsilon_g$ > 3.08 eV), a metal-insulator transition occurs upon the FM transition under strong correlation (large U). This is because double occupancy is completely suppressed and spin polarization leads to the complete filling of the lower spin subband. That is, within this specific region, the FM transition is accompanied by a Mott insulator transition.

To investigate the stability of the FM phase, we systematically analyze the variation of the PM-FM energy difference $\Delta$E with $\varepsilon_g$ for different U (Fig. 4d). Overall, $\Delta$E increases with decreasing $\varepsilon_g$ in the range $\varepsilon_g$ = 4.3 ~ 1.7 eV, reaching a peak near $\varepsilon_g$ ≈ 1.7 eV, which is similar to the ν = 2 case. The key distinction is that subsequently, within the range $\varepsilon_g$ = 1.7 ~ 0.0 eV, $\Delta$E continuously decreases and approaches zero. This is primarily due to the combined effects of the lower filling factor and the significant electron transfer resulting from p-d band overlap. Furthermore, as U decreases below 2.5 eV, this peak shifts toward smaller $\varepsilon_g$.

From the above results, particularly the spectral evolution with U shown in Appendix Fig. S1-S4, it can be seen that the dominant role of electron correlation U has a dual nature. First, under general conditions, U reduces the transition probability involving the *d*-orbitals, leading to orbital narrowing of the *d* states and a weakening of the *p-d* hybridization strength, thereby compressing the band gap—this is the common correlation effect observed in PM Kondo insulators [46]. Second, in the strong correlation limit (U = 10 eV), the situation undergoes a fundamental change. When double occupancy is completely suppressed ($b \to 0$) and the system achieves spin polarization ($p_- \to 0$), the renormalization factor for the occupied spin subband (+) satisfies:

$$\bar{z}_+ = (e^2 + p_-^2)^{-1/2}(ep_+ + p_-b)(b^2 + p_+^2)^{-1/2} \to 1. \qquad (17)$$

This limiting behavior is crucial, as it implies that the effective transition strength in this spin channel is no longer suppressed by U. The underlying reason is that, with the absence of opposite-spin electrons on the *d*-orbitals, the Hubbard repulsion—which normally hinders electron transitions by penalizing double occupancy—becomes irrelevant for the intra-spin-channel processes. Consequently, the band narrowing typically induced by correlations is circumvented. This provides the microscopic foundation for understanding the potentially large band gap that may be sustained in the Chern-Kondo insulator discussed later. Finally, throughout this process, the influence of U remains primarily confined to the energy level position and spin splitting of the *d*-orbitals, while its direct effect on the *p*-orbitals is weak—only exerting a limited perturbation through hybridization in the overlap region. Consequently, U alone is insufficient for achieving direct and effective control over the topological bands of the *p*-orbitals.

It is noteworthy that in the strong correlation limit (U = 10 eV), the system consistently exhibits complete spin polarization of the *d*-orbitals accompanied by the formation of a large local magnetic moment. From the perspective of band structure, since double occupancy is essentially fully suppressed, the energy level of the occupied *d*-orbital shifts significantly downward compared to its pre-polarization position (approaching the $\varepsilon_d$ value at U = 0), while a drastic splitting occurs between the spin subbands. Particularly for the ν = 2 case, after the spin subband splitting, electrons transfer from the *d*-orbitals to the *p*-orbitals, causing the dominant orbital character near the Fermi level to shift from *d* to *p*, while the *d*-orbital level sinks to a position far from the Fermi level. This series of transformations indicates that in the large-U limit, the system meets the Kondo limit conditions of the Anderson model—namely, the existence of a stable, large local moment and a Fermi level far removed from the *d*-orbital energy level—thus entering the physical regime described by the Kondo lattice model. Therefore, to explore substantive control of magnetism over the topological bands of *p*-orbitals, we will further introduce the s-d exchange interaction (i.e., the exchange coupling between local *d*-electron moments and itinerant *p*-electrons) on the basis of the aforementioned large-U limit (U = 10 eV) and investigate the case of J $\neq$ 0.

**C. Effects of the exchange interaction J in the strong correlation limit (U = 10 eV)**

(a)
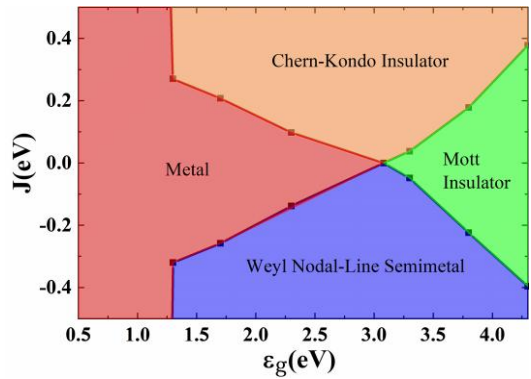

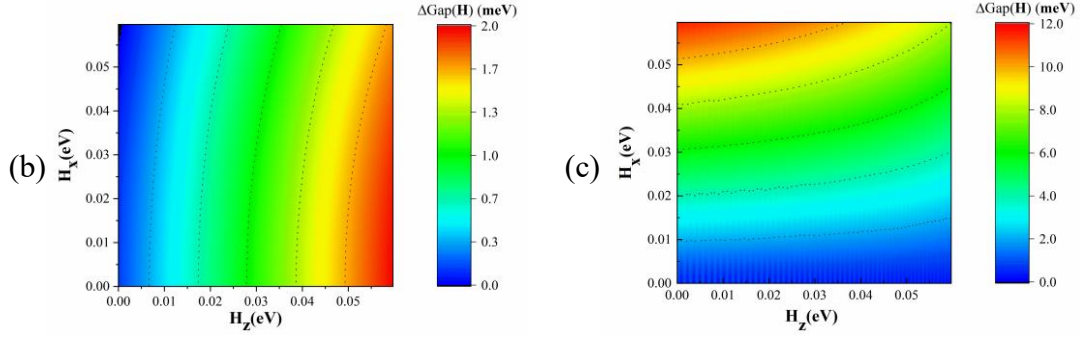

**Fig. 5. Topological phase diagram and magnetic field response of the system at filling ν = 1 in the strong correlation limit (U = 10 eV).**

(a) Topological phase diagram in the J-$\varepsilon_g$ parameter space. The background magnetic order is the FSP state. The red, green, orange, and purple regions correspond to the metallic phase, Mott insulator phase, Chern-Kondo insulator phase, and Weyl nodal-line semimetal phase, respectively. (b), (c) Magnetic field modulation of the topological band gap: shown are the changes in the band gap, defined as ΔGap(H) = Gap(H) − Gap(0), as a function of the transverse ($H_x$) and longitudinal ($H_z$) magnetic fields for (b) the Chern-Kondo insulator phase (J = 0.1 eV) and (c) the Weyl nodal-line semimetal phase (J = −0.1 eV), respectively. The calculations are performed for $\varepsilon_g$ = 3.3 eV.

First, it should be clarified that in our model, J > 0 corresponds to ferromagnetic exchange coupling between the local *d*-electron moments and the itinerant *p*-electrons, while J < 0 represents antiferromagnetic exchange coupling.

For the system with ν = 2, as shown by the representative spectra in Appendix Fig. S5, the Fermi level vicinity is primarily dominated by *p*-orbitals and the system is metallic in the strong correlation limit. Here, introducing a finite J is equivalent to applying an effective exchange field to the *p*-orbitals, whose main effect is to induce a Zeeman splitting in these orbitals, while exerting negligible influence on the topology near the Fermi level. Although changes in band topology can be observed in deep energy levels far from the Fermi energy, their contribution to low-energy physics is limited. This result suggests that the pronounced topological modulating effects of the exchange interaction J are more likely to occur for the filling case of ν = 1; therefore, the following discussion will focus on this scenario.

For the system with ν = 1, Appendix Fig. S6 displays the evolution of the energy spectra with J in the strong correlation limit for typical values of $\varepsilon_g$. Taking the system with $\varepsilon_g$ = 3.3 eV as an example: at J = 0, it is a trivial Mott insulator. Introducing a finite J effectively induces a *p-d* band inversion, yet the topological outcome depends critically on the sign of J.

When J > 0, the system opens a topologically nontrivial gap, forming a topological state with a Chern number of 1. This is a strongly correlated topological state, distinct from the QAHE state in a single-particle picture, wherein the *d*-orbitals simultaneously contribute to both magnetic order and nontrivial topology. Such a state is referred to in theoretical studies as a "Chern-Kondo insulator" and has been realized in optical lattice systems [49]. When J < 0, although a global gap does not open, the p-d band crossing forms a closed nodal loop. Since this degeneracy is two-fold, we term this state a Weyl-type nodal-line semimetal. The reason why the sign of J leads to such distinct topological states is as follows: To simplify the model, we have included SOC only in the on-site energy part and neglected the spin-flip processes in the s-d exchange interaction. This results in the complete decoupling of the spin-up and spin-down subspaces of the system, leading to an intrinsic asymmetry in how positive and negative J regulate the bands.

For the system with $\varepsilon_g$ = 2.3 eV, although it is metallic at J = 0 due to *p-d* band overlap, adjusting J can still effectively shift the *p*-orbital bands, thereby enabling topological phase transitions analogous to the $\varepsilon_g$ = 3.3 eV case—namely, inducing either a Chern-Kondo insulator or a Weyl nodal-line semimetal.

To systematically characterize the phase transitions in the strong correlation limit, we plot the J-$\varepsilon_g$ phase diagram in Fig. 5(a). Since the system remains in the FSP phase throughout this parameter region, the diagram primarily reveals the evolution of its topological properties. It can be seen that when J > 0, the system can enter the Chern-Kondo insulator phase, whereas when J < 0, it transitions into the Weyl-type nodal-line semimetal phase. Notably, the critical |J| value required to induce the topological transition depends strongly on $\varepsilon_g$: once $\varepsilon_g$ deviates from the band inversion critical point ($\varepsilon_g \approx 3.08$ eV), the required |J| increases rapidly. Particularly in the strong orbital overlap region ($\varepsilon_g$ < 1.7 eV), the critical |J| rises sharply, and for $\varepsilon_g$ < 1.3 eV, the system remains metallic for any finite |J|. In contrast, in the insulating region without orbital overlap ($\varepsilon_g$ > 3.08 eV), J consistently exhibits a significant modulating capability over the band topology.

It is important to note that Weyl points in three dimensions possess topological protection stability, whereas the Weyl nodal lines appearing for J < 0 in our two-dimensional system are inherently dynamically unstable [56]. These nodal lines originate from *p-d* band crossings between different spin channels. Considering additional realistic perturbations—such as non-local spin-orbit coupling not included in the current model, or applying an in-plane transverse magnetic field—could potentially gap out the nodal lines, possibly inducing a Chern-Kondo insulator.

On the other hand, the Chern-Kondo insulator induced by J > 0 stems from *p-d* band inversion within the same spin channel. In contrast, the Weyl nodal-line semimetal caused by J < 0 involves band crossings between different spin channels. This fundamental difference in origin leads to their markedly distinct responses to an external magnetic field.

To reveal this difference, we introduce a Zeeman coupling term of the following form:

$$H_B = -\sum_{\mathbf{k},\sigma,\sigma'} \hat{\phi}^\dagger_{\mathbf{k}\sigma} [\mathbf{H} \cdot \hat{\boldsymbol{\tau}}_{\sigma\sigma'} \otimes \hat{I}] \hat{\phi}_{\mathbf{k}\sigma'} \tag{18}$$

Here, the magnetic field vector **H** incorporates the Bohr magneton and the g-factor, its magnitude directly representing the effective Zeeman energy. We calculated the variation of the band gap with the **H** for the system with $\varepsilon_g$ = 3.3 eV at J = ±0.1 eV; the results are shown in Figs. 5(b) and 5(c).

The topological gap of the Chern-Kondo insulator at J = 0.1 eV is insensitive to both transverse ($H_x$) and longitudinal ($H_z$) magnetic fields, demonstrating robust stability. In stark contrast, for the Weyl nodal-line semimetal at J = −0.1 eV, the degenerate nodal lines can be gapped out under a transverse magnetic field ($H_x$), giving rise to a magnetic gap that increases monotonically with $|H_x|$. This specific, direction-sensitive response to the magnetic field is in excellent agreement with recent experimental observations in multilayer $MnBi_2Te_4$ thin films [57].

## VI. SUMMARY

In this work, we systematically investigate the regulatory mechanism of electron correlation strength U on the band structure of magnetic topological materials by constructing and studying a low-energy effective model that combines strong correlation and topology. Our study focuses on paramagnetic (PM) and ferromagnetic (FM) orders, motivated by the material foundation of monolayer $MnBi_2Te_4$ and its analogues, in which intra-layer magnetic coupling is intrinsically FM due to near-90° Mn–Te–Mn superexchange, as supported by the Goodenough-Kanamori rules and experimental observations.

Our results demonstrate that increasing U can effectively drive the system from a PM phase into a FM phase, but the specific pathway is closely related to the electron filling factor $\nu$. For systems with $\nu = 2$ that are trivial insulators at $U = 0$, enhancing U in the PM phase tunes the *d*-orbital energy levels, promotes *p-d* band overlap and inversion, and realizes a transition from a trivial insulator to a topological insulator, exhibiting behavior typical of a Kondo topological insulator. Notably, based on the three-band model employed in this study, a further increase in $U$ from this topological insulating state drives the highest-energy third band to cross the Fermi level. This metallizes the system and facilitates the completion of the PM-to-FM phase transition. For metallic systems at half-filling ($\nu = 1$), increasing U also induces a PM-FM transition. However, it manifests as a direct jump from the PM phase to a FSP phase, bypassing any intermediate gradual FM state. Furthermore, for $\nu = 1$ systems with no initial *p-d* band overlap, the FM transition is accompanied by a metal-insulator transition, leading to the formation of a Mott insulator.

This study finds that the Hubbard-type correlation U primarily regulates the energy level and spin splitting of the *d*-orbitals, with a limited effect on the *p*-orbitals. However, in the large-U limit, the system enters an FSP phase where double occupancy is suppressed, and its physics transitions from the regime of the Anderson model to that of the Kondo model. Here, the formed large local moments can effectively modulate the itinerant p electrons via the s-d exchange interaction (J). In systems with $\nu = 1$, by introducing the exchange interaction J, we successfully realize topological states such as the Chern-Kondo insulator and the two-dimensional Weyl-type nodal-line semimetal. The magnetic gap opened in the latter under a transverse magnetic field exhibits a highly sensitive and specific response, in excellent agreement with recent experimental observations of the QAHE enhancement by a transverse field in $MnBi_2Te_4$ systems [57].

Regarding magnetic control, calculations show that appropriately reducing the *p-d*-orbital energy difference $\varepsilon_g$ at $U = 0$—which introduces moderate band overlap near the Fermi level—helps enhance the PM-FM energy difference, thereby improving the stability of the FM phase. It also lowers the critical $U_p$ value required to achieve spin polarization while only weakly suppressing the saturated magnetic moment. Experimentally, this can be analogous to tuning orbital on-site energies via gate voltage. Particularly for systems where the *d*-orbital resides at an inversion center, an external electric field can cause a relative energy offset between the upper and lower *p*-orbital layers. This selectively enhances the hybridization between the *d*-orbital and one of the

*p*-orbital layers, enabling more effective electrical control of magnetism. Furthermore, recent studies have integrated MnBi$_2$Te$_4$ with layered ferroelectric materials In$_2$Te$_3$ into van der Waals heterostructures [58]. By leveraging the non-volatile polarization of the ferroelectric layer, a strong and switchable interfacial electric field is generated. This field can effectively manipulate the band structure and magnetic order of the adjacent MnBi$_2$Te$_4$, and even induce topological phase transitions, thereby offering a more reliable control approach compared to gate-voltage techniques.

Concerning the control of topological properties, this study reveals the following core principles and prospects:

**(i).** Systematic evolution of the band structure under correlation: At $U = 0$, the vicinity of the Fermi level is dominated by *d*-orbitals; entering the weak correlation regime (small U), the *d*-orbitals undergo slight spin splitting but retain their dominance; when U increases to the strong correlation limit, double occupancy is completely suppressed, and the *d*-orbitals experience complete spin splitting and essentially sink below the Fermi level. At this point, if the s-d exchange interaction between local moments and itinerant electrons is further considered, the originally lower-energy *p*-orbitals—which lie below the *d*-orbitals and are not included in our simplified model—will be significantly lifted. Consequently, the low-energy region near the Fermi level becomes dominated by *p*-orbitals, with *d*-orbitals relegated to a secondary role. This evolution picture is consistent with the band features revealed by DFT calculations for MnBi$_2$Te$_4$ under correlation. Within this framework, the system enters the regime described by the Kondo lattice model. Here, magnetism modulates the *p*-orbital bands via the s-d exchange interaction; however, the emergence of possible nontrivial topological states often requires the inclusion of *p-p* orbital interactions beyond the scope of our current model. This extends past the framework of acquiring topology solely through *p-d* hybridization as described in the present work.

**(ii)** We propose and demonstrate a class of "Chern-Kondo insulators" with a potentially large band gap. This state is characterized by strongly correlated *d*-orbitals that simultaneously contribute to both magnetic order and nontrivial topology. Its key physical foundation is that the *d*-orbitals are in an almost fully polarized state. Unlike conventional PM Kondo insulators [46]—where correlation effects induce significant band renormalization and weaken effective hybridization, leading to a small topological gap—in the Chern-Kondo insulator, the combination of strong suppression of double occupancy and *d*-orbital spin polarization ensures that correlation effects do not cause band narrowing nor weaken the *p-d* hybridization strength. Consequently, this state is expected to possess a larger topological gap compared to traditional Kondo insulators, offering a novel route for observing the QAHE at elevated temperatures.

The key to realizing this state lies in hole doping to tune the Fermi level of the system to the vicinity of the $\nu = 1$ filling as specified by our model. Feasible experimental pathways include: employing ionic liquid gating [59,60] for in situ carrier modulation in Kondo lattice materials with analogous electronic structures (such as Ce- [61,62] or Yb-based [63,64] compounds) or in artificial heterostructures; alternatively, introducing holes via $Ca^{2+}$ doping to replace $Sm^{3+}$ in $SmB_6$ [65], a material confirmed as a candidate topological Kondo insulator. These approaches provide practical and viable experimental schemes for verifying our theoretical prediction.

## DATA AVAILABILITY
All data generated or analysed during this study are included in this published article.

## AUTHOR CONTRIBUTIONS
Z.-Y. Wang developed the model, performed calculations, and wrote the manuscript; Y.-M. Quan optimized the computational codes and contributed to discussions; Y.-X. Sun performed DFT calculations; L.-J. Zou and X.-L. Yu supervised the project, acquired funding, and revised the manuscript.

## ACKNOWLEDGMENT
X.-L. Yu acknowledges support from the Shenzhen Science and Technology Program (Grant No. JCYJ20250604174400001), the Basic Start-up Fund for Introduced Talents at Sun Yat-sen University and the Guangdong Basic and Applied Basic Research Foundation (Grant No. 2023A1515011852).
L.-J. Zou acknowledges support from the National Natural Science Foundation of China (Grant No. 11534010).

## COMPETING INTERESTS
The authors declare no competing interests.

## APPENDIX
Throughout this appendix, red and blue coloring in the spectra indicates dominant *d*-

orbital and *p*-orbital character, respectively.

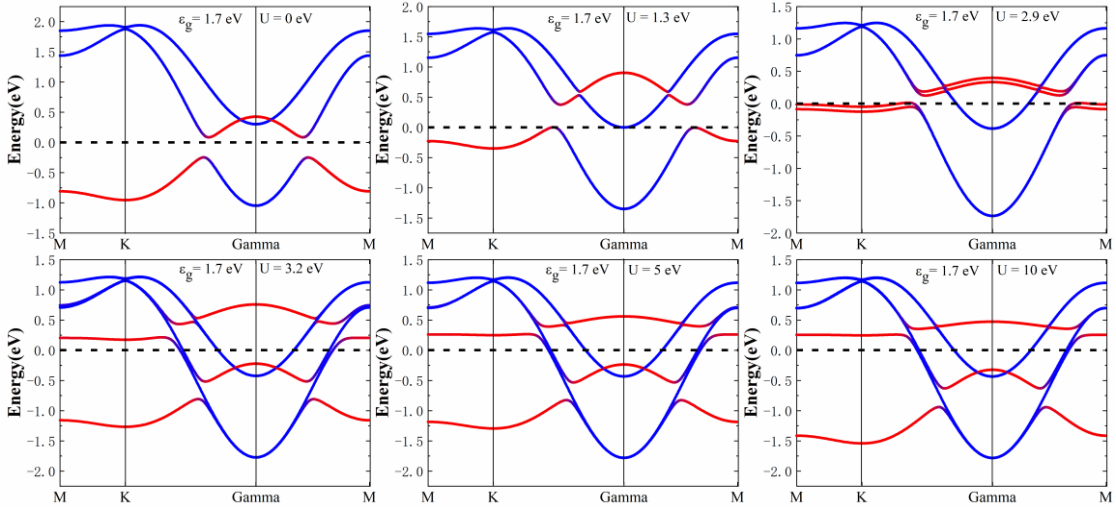

**Fig. S1. Spectral evolution with U (0 to 10 eV) for the system with ν = 2 and $\varepsilon_g$ = 1.7 eV.** Here, U = 1.3 eV and U = 2.9 eV correspond to the critical points before the PM-FM and FM-FSP phase transitions, respectively. It is noteworthy that the spectrum at U = 1.3 eV lies within the critical region of the insulator-metal crossover. For U > 3 eV (after spin polarization), the band structure below the Fermi level remains largely unchanged.

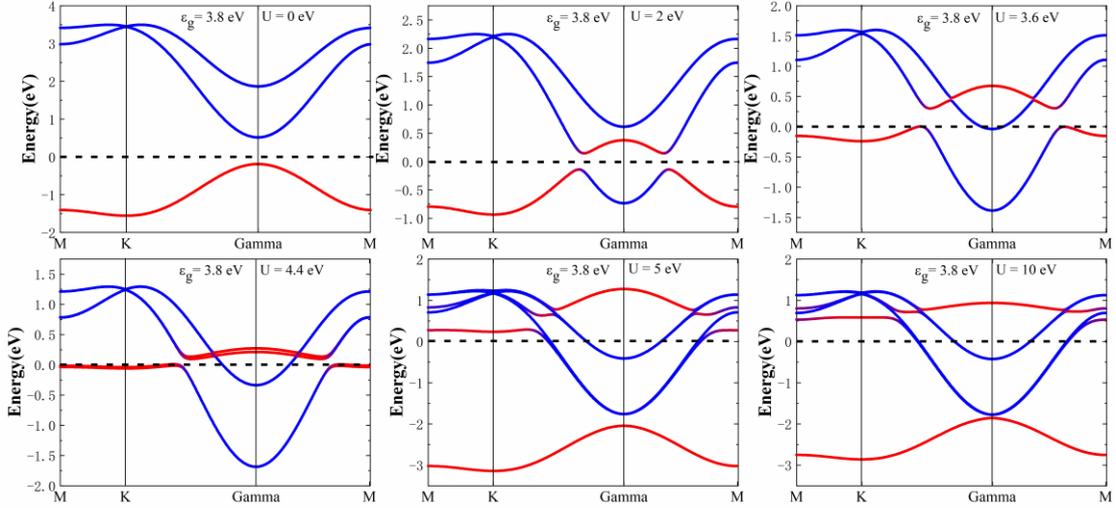

**Fig. S2. Spectral evolution with U (0 to 10 eV) for the system with ν = 2 and $\varepsilon_g$ = 3.8 eV.** As U increases from 0 to 2 eV, the system undergoes a transition from a trivial insulator to a topological insulator. U = 3.6 eV and U = 4.4 eV correspond to the critical points before the PM-FM and FM-FSP phase transitions, respectively, with the spectrum at U = 3.6 eV also residing in the critical region of the insulator-metal crossover. For U > 4.5 eV (after spin polarization), the band structure below the Fermi level remains largely unchanged.

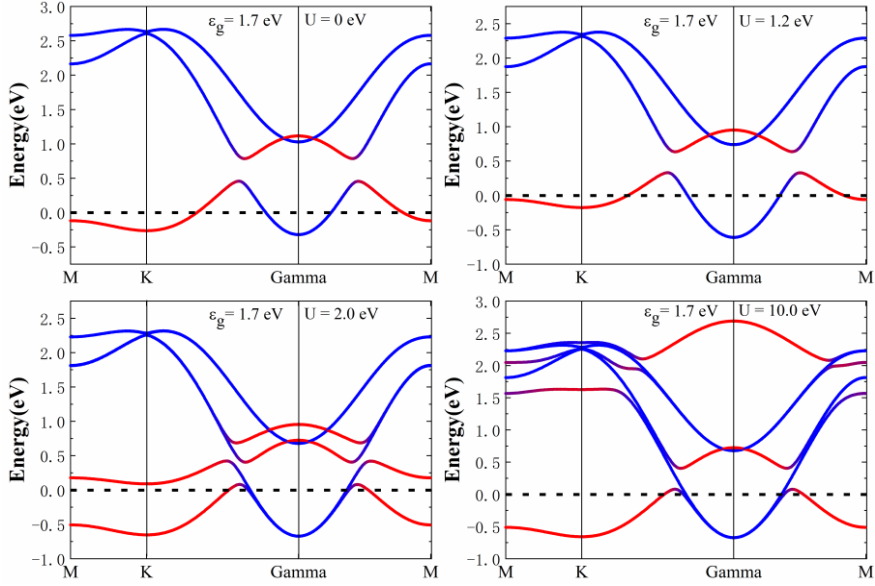

**Fig. S3. Spectral evolution with U (0 to 10 eV) for the system with ν = 1 and $\varepsilon_g$ = 1.7 eV.** U = 1.2 eV marks the critical point before the PM-FSP phase transition. Thereafter, the band structure below the Fermi level remains largely unchanged.

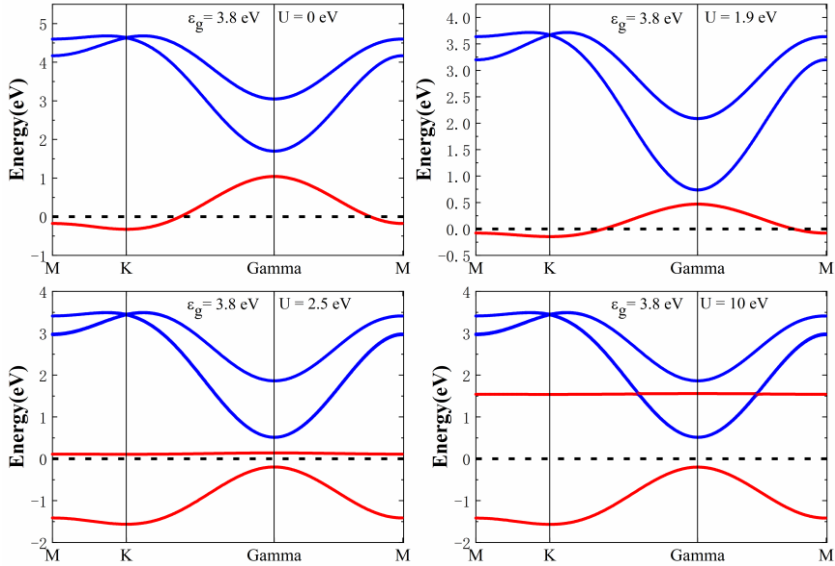

**Fig. S4. Spectral evolution with U (0 to 10 eV) for the system with ν = 1 and $\varepsilon_g$ = 3.8 eV.** U = 1.9 eV marks the critical point before the PM-FSP phase transition. Thereafter, the band structure below the Fermi level remains largely unchanged.

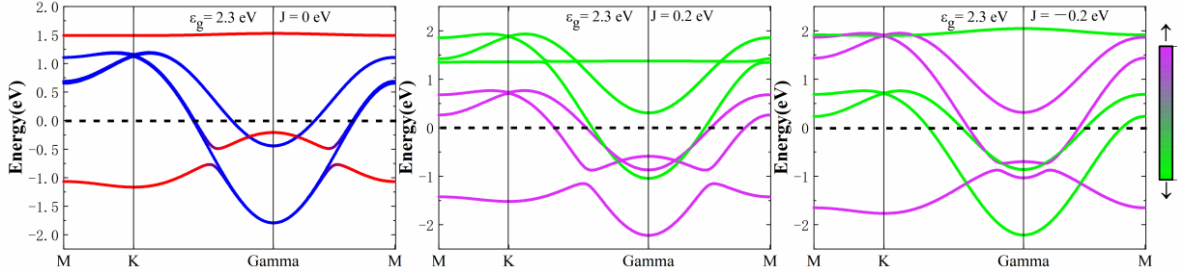

**Fig. S5. Band spectra for the system with ν = 2 and $\varepsilon_g$ = 2.3 eV at J = 0, ±0.2 eV in the strong correlation limit (U = 10 eV).** Purple and blue segments denote the spin-up and spin-down channels, respectively.

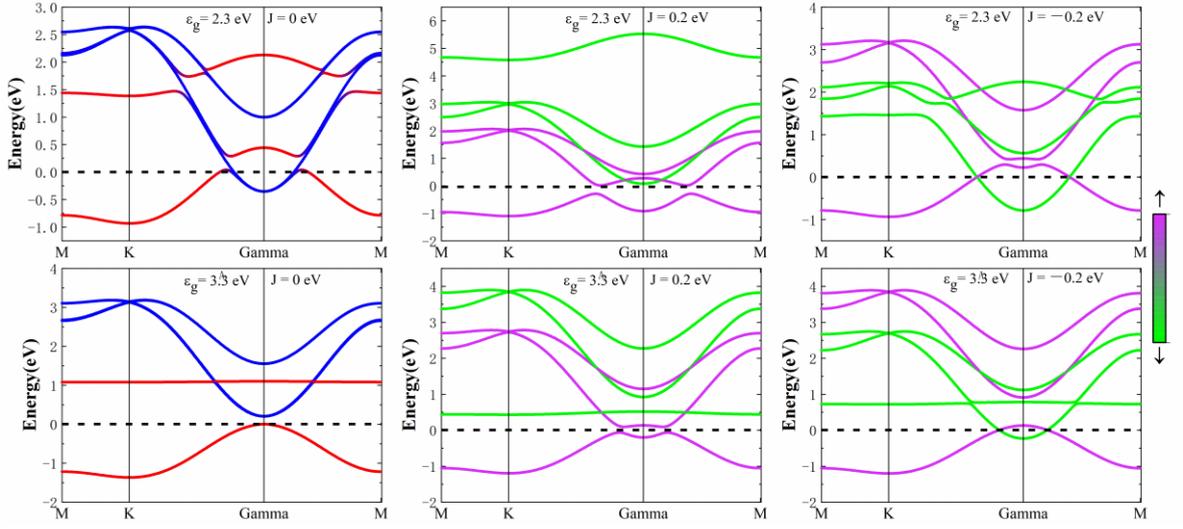

**Fig. S6. Band spectra for systems with ν = 1 at $\varepsilon_g$ = 2.3 eV and $\varepsilon_g$ = 3.3 eV for J = 0, ±0.2 eV in the strong correlation limit (U = 10 eV).** Purple and blue segments denote the spin-up and spin-down channels, respectively.